\documentclass[12pt]{aastex}

\usepackage{newtxtext,newtxmath}
\usepackage{graphicx}	% Including figure files
\usepackage{amsmath}	% Advanced maths commands
\usepackage{amssymb}	% Extra maths symbols
\usepackage{color}
\usepackage{bm}%bold matH
\usepackage{epsf}
\usepackage{wrapfig}
\usepackage{multirow}
\usepackage{epstopdf}
\usepackage{natbib}

\begin{document}
%%%%%%%%%%%%%%%%%%% TITLE PAGE %%%%%%%%%%%%%%%%%%%

% Title of the paper, and the short title which is used in the headers.
% Keep the title short and informative.
\title{Hall Effect in the coma of 67P/Churyumov-Gerasimenko}

% The list of authors, and the short list which is used in the headers.
% If you need two or more lines of authors, add an extra line using \newauthor
\author{Z. Huang\altaffilmark{1}, 
G. T\'oth\altaffilmark{1}, 
T. I. Gombosi\altaffilmark{1}, 
X. Jia\altaffilmark{1}, 
M. R. Combi\altaffilmark{1}, 
K. C. Hansen\altaffilmark{1}, 
N. Fougere\altaffilmark{1}, \\
Y. Shou\altaffilmark{1}, 
V. Tenishev\altaffilmark{1}, 
K. Altwegg\altaffilmark{2}
and M. Rubin\altaffilmark{2}}

% List of institutions
\altaffiltext{1}{\emph{Climate and Space Sciences and Engineering, University of Michigan, Ann Arbor, MI 48109, USA}}
\altaffiltext{2}{\emph{Physikalisches Institut, University of Bern, Bern, Switzerland}}

\begin{abstract}
Magnetohydrodynamics simulations have been carried out in studying the solar wind and cometary plasma interactions 
for decades. Various plasma boundaries have been simulated and compared well with observations
for comet 1P/Halley. The Rosetta mission, which studies comet 67P/Churyumov-Gerasimenko,
challenges our understanding of the solar wind and comet interactions. The Rosetta Plasma Consortium
observed regions of very weak magnetic field outside the predicted diamagnetic cavity. In this paper, we simulate the inner coma with the
Hall magnetohydrodynamics equations and show that the Hall effect is important in the inner coma environment.
The magnetic field topology becomes complex and magnetic reconnection occurs on the dayside when the Hall effect is taken
into account. The magnetic reconnection on the dayside can generate weak magnetic filed regions outside the global
diamagnetic cavity, which may explain the Rosetta Plasma Consortium observations. 
We conclude that the substantial change in the inner coma environment is due to the fact that
the ion inertial length (or gyro radius) is not much smaller than the size of the diamagnetic cavity.

\end{abstract}


\clearpage

%%%%%%%%%%%%%%%%%%%%%%%%%%%%%%%%%%%%%%%%%%%%%%%%%%

%%%%%%%%%%%%%%%%% BODY OF PAPER %%%%%%%%%%%%%%%%%%

\section{Introduction}
Cometary magnetospheres is one of the most important topics in planetary science. Because the nucleus of a comet is usually very small in size ranging
from a few hundred meters to tens of kilometers (e.g., the radius of the nucleus for comet 1P/Halley is about 10\,km) and the gravity is extremely weak
(usually considered negligible when simulating the cometary neutral gas and plasma), the cometary coma is much larger in size compared to the nucleus itself.
For example, Giotto observed plasma boundaries of comet 1P/Halley starting at roughly 1\,Mkm away from the nucleus \citep{Reme_1986}.
The cometary magnetosphere resulting from the solar wind interaction with the coma has some distinct features from the magnetospheres associated with
planets or planetary moons, such as the formation of a diamagnetic cavity.
\cite{Gombosi_2015} provided an excellent review of the cometary magnetosphere.
A typical cometary magnetosphere for an active comet near perihelion includes
a bow shock which slows down the supersonic solar wind to subsonic speed \citep{Galeev_1985, Koenders_2013}, 
a diamagnetic cavity inside which the magnetic field drops to zero \citep{Neubauer_1986, Cravens_1986, Goetz_2016, Goetz_2016_2}, 
a recombination layer which separates the inner shock (which slows down the supersonic cometary ion outflow to subsonic) and
the contact surface (where the solar wind protons cannot penetrate). One of the primary goals of the 
Rosetta Plasma Consortium (RPC) was to observe the evolution of the solar wind and comet interactions.
However, due to the close proximity of the Rosetta spacecraft to the nucleus, RPC was not able to observe the bow shock.
Also the recombination layer and contact surface have not been clearly identified to date.
Based on RPC observations, \cite{Mandt_2016} reported plasma boundaries separating an inner region and an outer region and they
concluded the observed plasma boundaries are an ion-neutral collisionopause boundary, 
which has not been predicted by previous numerical simulations \citep{Rubin_2015, Koenders_2015, Huang_2016}. 
In addition to this unpredicted boundary, the magnetic field observed by RPC is also unexpected:
the diamagnetic `cavity' \citep{Goetz_2016, Goetz_2016_2} was observed much farther away than the 
predicted locations \citep{Rubin_2015, Koenders_2015, Huang_2016}. 

Lots of effort has been made in numerical simulations to understand the solar wind
and comet interactions. There are two major approaches in simulating the cometary environment:
the fluid approach \citep{Gombosi_1996, Hansen_2007, Rubin_2014, Rubin_2015, Huang_2016} 
and the hybrid approach \citep{Bagdonat_2002, Koenders_2015,Wedlund_2017}. In a fluid approach, 
the plasma is treated as fluids and governed by the magnetohydrodynamic (MHD) equations. 
The fluid approach is an accurate description of the 
macroscopic quantities of the plasma 
when the Knudsen number ($ \text{K}_\text{n} = \frac{\lambda}{l}$, where
$\lambda$ is the mean free path and $l$ is the characteristic length scale of the flow) is much smaller
than unity. On the other hand, the hybrid approach simulates ions as individual particles and electrons as a fluid.
The hybrid approach can capture the kinetic features of the plasma and works well also for large Knudsen numbers.
Compared to the fluid approach, the hybrid approach is very computationally expensive and 
is usually limited to a small simulation domain. 
Recently, a third approach to simulate the solar wind and comet interactions was developed by \cite{Deca_2017} 
with a fully kinetic code, which treats both ions and electrons as particles. They simulated 
comet CG at 3\,au and showed that their simulations agreed well with RPC observations at that heliocentric distance. 
However, their code does not include any chemical reactions and collisions between particles, which makes it 
not applicable to comets near perihelion, where chemical reactions are important.
Also their model is limited to a small domain that does not include the bow shock.

There have been extensive discussions about whether fluid codes can properly simulate cometary environments 
or those of other planets/moons without an intrinsic magnetic field, in which case the ion gyro radius is larger than the 
length scale of interaction regions. We argue that fluid codes are not limited exclusively by the requirement that the length scale 
must be much larger than the ion gyro radius, because collisions among different particles as well as chemical 
reactions may reduce the ion kinetic effects arising from the gyro motion. It is true that single fluid ideal MHD models 
cannot capture any ion gyration effects compared to a hybrid simulation \citep{Hansen_2007}. But when fluid 
simulations take into account different fluids with different velocities as well as collisions among them, multi-fluid 
MHD models are capable of capturing some important ion kinetic behaviors. For example, \cite{Rubin_2014} 
showed that their multi-fluid MHD model is capable of resolving the gyration of different ion fluids with reasonably 
good agreement with what has been predicted by a hybrid model \citep{Muller_2011}. The major discrepancy lies 
in the very inner coma region with striations/filaments in the cometary ion density \citep{Koenders_2015}. 
On the other hand, multi-fluid simulations for other planets/moons without an intrinsic magnetic field have demonstrated 
that the simulation results agree well with in-situ plasma observations \citep{Najib_2011, Bougher_2015, Ma_2011}. 
Hybrid models certainly provide a better description of the plasma environment near the comet nucleus. 
However, it is much more expensive or nearly impossible computationally to run hybrid models with a grid 
resolution comparable to those of fluid models on a large enough domain to properly set up the outer boundary 
conditions and to resolve the details of a diamagnetic cavity formed close to the nucleus. With a relatively coarse 
grid typically used in hybrid models, effects of numerical diffusion are expected to be much stronger than in fluid 
models and in such a case, the evolution of the magnetic field may not be properly described. As described below, 
the multi-fluid Hall MHD model presented in this paper represents another step in further resolving kinetic effects with fluid simulations.

One of the imperfections of previous MHD models applied in cometary studies is that the 
Hall effect is usually not taken into account. The Hall effect describes the relative
speed (current) between ions and electrons and appears in the generalized Ohm's law. 
This current may affect the magnetic field evolution in the system if the Hall effect is taken into account
in the induction equation. The Hall effect is important in magnetic 
reconnection studies as Hall MHD is the minimal modification of resistive MHD that can reproduce the fast
reconnection process \citep{Birn_2001}, partially due to the strong current near the reconnection null point.
In the cometary magnetosphere, the diamagnetic cavity is a unique feature that other planets/moons do not
have. As the magnetic field drops to zero in a short distance, there must be strong currents
along the diamagnetic cavity boundary. How these currents affect the inner coma environment is
still unknown. In this paper, we simulate the inner coma environment with Hall MHD equations and
show that the Hall effect is important in the inner coma and
 the classical plasma boundaries obtained by previous models need to be revisited.
The detailed model description can be found in the appendix.

\section{The Hall MHD model}
Our Hall MHD model is an extension of the multifluid
model developed by \cite{Huang_2016}.
In the following equations, mass density, velocity vector, pressure, the identity matrix and the adiabatic index
are denoted by symbols $\rho$, $\mathbf{u}$, $p$, ${I}$ and $\gamma$, respectively. 
The cometary neutral gas, the ions (cometary and solar wind) and the electrons are denoted by subscripts $n$, $s$ and $e$, respectively.
The symbol $Z$ denotes the ion charge state while the symbol $e$ is for the unit charge.

There are four fluids in the model. One fluid describes the cometary neutral gas with the Euler equations:
\begin{subequations}
\begin{gather}
\frac{\partial \rho_n}{\partial t} + \nabla \cdot (\rho_n \mathbf{u_n}) = \frac{\delta \rho_n}{\delta t} 
%\label{eqn:hd-a}
\\
\frac{\partial \rho_n \mathbf{u_n}}{\partial t} + \nabla \cdot (\rho_n \mathbf{u_n}\mathbf{u_n} + p_n {I} ) = \frac{\delta \rho_n \mathbf{u_n}}{\delta t} 
%\label{eqn:hd-b}
\\
\frac{\partial p_n}{\partial t}  + \nabla \cdot ( p_n   \mathbf{u_n}) + (\gamma_n - 1)p_n (\nabla\cdot\mathbf{u_n}) = \frac{\delta p_n}{\delta t} 
%\label{eqn:hd-c}
\end{gather}
\label{eqn:hd}
\end{subequations}
and the other two fluids describe the cometary ions and the solar wind protons with the multifluid MHD equations, which are solved individually for both fluids:
\begin{subequations}
\begin{gather}
\frac{\partial \rho_s}{\partial t} + \nabla \cdot (\rho_s \mathbf{u_s}) = \frac{\delta \rho_s}{\delta t} 
%\label{eqn:mhd-a}
\\
\begin{split}
\frac{\partial \rho_s \mathbf{u_s}}{\partial t} & + \nabla \cdot (\rho_s \mathbf{u_s}\mathbf{u_s} + p_s \text{I} )
\\&
     - Z_se\frac{\rho_s}{m_s}(\mathbf{E+\mathbf{u_s}\times\mathbf{B}})= \frac{\delta \rho_s \mathbf{u_s}}{\delta t}
\end{split}
%\label{eqn:mhd-b}
\\
\frac{\partial p_s}{\partial t}  + \nabla \cdot ( p_s   \mathbf{u_s}) + (\gamma_s - 1)p_s (\nabla\cdot\mathbf{u_s}) = \frac{\delta p_s}{\delta t}
%\label{eqn:mhd-c}
%\label{eqn:mhd-d}
\end{gather}
\label{eqn:mhd}
\end{subequations}

For the electrons, we do not specify the continuity and momentum equations.
Assuming charge neutrality in the plasma, the electron number density can be obtained as $n_e = \sum_{s=ions} Z_s n_s$.
The electron velocity $\bf u_e$ is obtained from $\mathbf{u_e} = \mathbf{u_+} + \mathbf{u_H}$, where $\bf u_+$ is the 
charge averaged ion velocity ($\mathbf{u_+} = \frac{\sum_{s=ions} Z_s n_s \mathbf{u_s}}{n_e}$)  
and $\bf u_H$ is the Hall velocity ($ \mathbf{u_H} = - \frac{\mathbf{j}}{n_e e}$, 
where $\bf j$ is the current density $\mathbf{j =(1/\mu_0) \nabla \times B}$).
The electron pressure in the system is described by Equation \ref{eqn:pe}:
\begin{equation}
\frac{\partial p_e}{\partial t}  + \nabla \cdot ( p_e   \mathbf{u_e}) + (\gamma_e - 1)p_e (\nabla\cdot\mathbf{u_e}) = \frac{\delta p_e}{\delta t}
\label{eqn:pe}
\end{equation}

We use Equations\,(\ref{eqn:hd})\,-\,(\ref{eqn:pe}) to describe the behavior and interactions of 
different fluids (the cometary neutral gas, the cometary ions, the solar wind protons, and the electrons) in the system. 
Ionization (photo-ionization and electron impact ionization) of the cometary neutral gas, charge exchange between
neutrals and ions, collisions (elastic and inelastic) between different fluids, and recombination are all taken into 
account in simulating the cometary environment and they appear as source terms 
in the right hand side of Equations\,(\ref{eqn:hd})\,-\,(\ref{eqn:pe}). We apply the same source terms as
\cite{Huang_2016}. The stiffness of the source terms may limit the time step, so a point-implicit algorithm 
\citep{Toth_2012} is applied to evaluate these terms.

The electric and magnetic fields are also needed to solve the multifluid equations. The electric field is derived from the electron momentum equation
if the inertial terms are assumed to be zero (due to the small electron mass):
\begin{equation}
\mathbf{E} = - \mathbf{u_e} \times \mathbf{B} - \frac{1}{n_e e}  \nabla  p_e
\label{eqn:efield}
\end{equation}
The magnetic field is obtained from the induction equation:
\begin{equation}
\frac{\partial \mathbf{B}}{\partial t} = - \nabla \times \mathbf{E}
\label{eqn:hall}
\end{equation}

We solve Equations (\ref{eqn:hd}) to (\ref{eqn:hall}) 
on a 3D block adaptive grid with the BATS-R-US (Block-Adaptive Tree Solarwind Roe-type Upwind Scheme) 
code \citep{Powell_1999, Toth_2012}. The beauty of the adaptive grid is that we can resolve different length scales
in the system, so that the simulation can resolve the nucleus while modeling the global scales. In the comet CG case,
the radius of the nucleus is about 2\,km, the global diamagnetic cavity is reported to be about 100\,km, and
the bow shock is expected to be at about 8,000\,km upstream of the nucleus \citep{Rubin_2015, Koenders_2015, Huang_2016}. In our simulation,
the smallest cell is located near the nucleus with the size of about $0.12\,\text{km}$ and the largest cell 
is located near the outer boundary with the size of about 31,250 km, which requires 18 levels of refinements 
(each refinement level increases the resolution by a factor of two) in the domain.
We use the Cometocentric Solar Equatorial\,(CSEQ) frame in the simulation. In this frame, +x
points toward the Sun, the z axis contains the solar rotation axis, and the y axis is orthogonal to the x and z axes. 
The solar wind is considered to move along the -x direction with the interplanetary magnetic field points in the +y direction
at the upstream boundary. 
The simulation box is within $\pm10^6\,\text{km}$ in the x direction 
and $\pm 0.5\times10^6\,\text{km}$ in both y and z directions.
We specify boundary conditions the same way as \cite{Huang_2016} at the edge of the 
simulation box (outer boundary) as well as at the nucleus surface (inner boundary).

In the present study, the cometary neutral gas is limited to water molecules with the specific heat ratio ($\gamma$) of
$\frac{4}{3}$ and the corresponding cometary ions are $\text{H}_2\text{O}^+$ with the same $\gamma$. 
$\gamma = \frac{5}{3}$ is applied for the solar wind protons as well as electrons.
An idealized spherical comet with the neutral gas outflow driven by the solar illumination (hereafter illuminated sphere) seems to be the minimum 
requirement not to lose important asymmetrical features in the inner coma \citep{Huang_2016},
so we apply this nucleus condition at the inner boundary, which is the same as Case 2 in \cite{Huang_2016}. 
We apply the same input parameters listed in Tables 2 and 3 in \cite{Huang_2016}.
We first run the multifluid model in steady state mode without the Hall effect to reach
a steady state. We then introduce the Hall effect at $t=0$ and run the model in time-dependent mode to investigate the evolution of the inner coma.

\section{Simulation results}

\begin{figure*}[ht!]
\begin{tabular}{cc}
\includegraphics[width=0.5\linewidth]{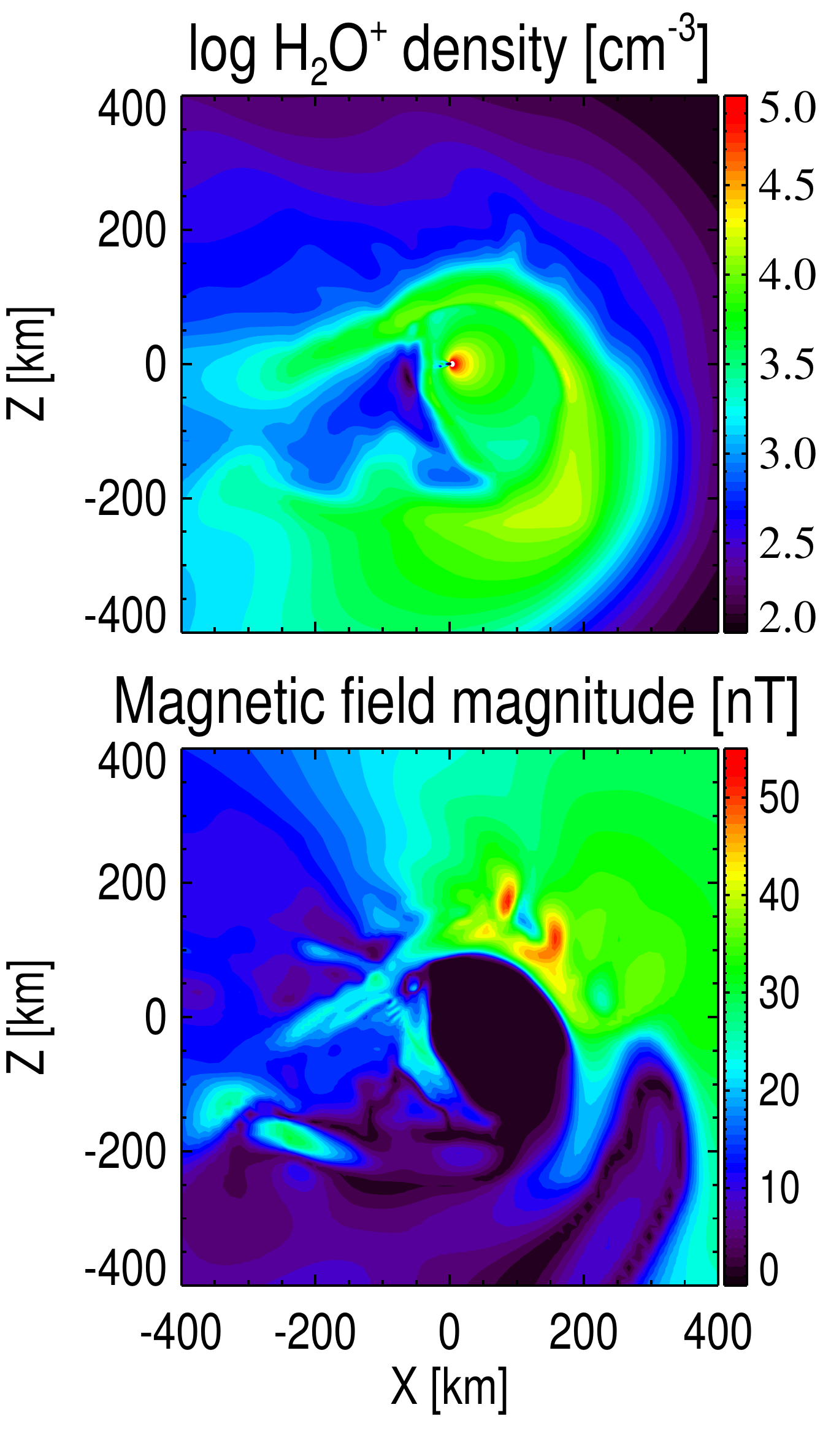} &
\includegraphics[width=0.5\linewidth]{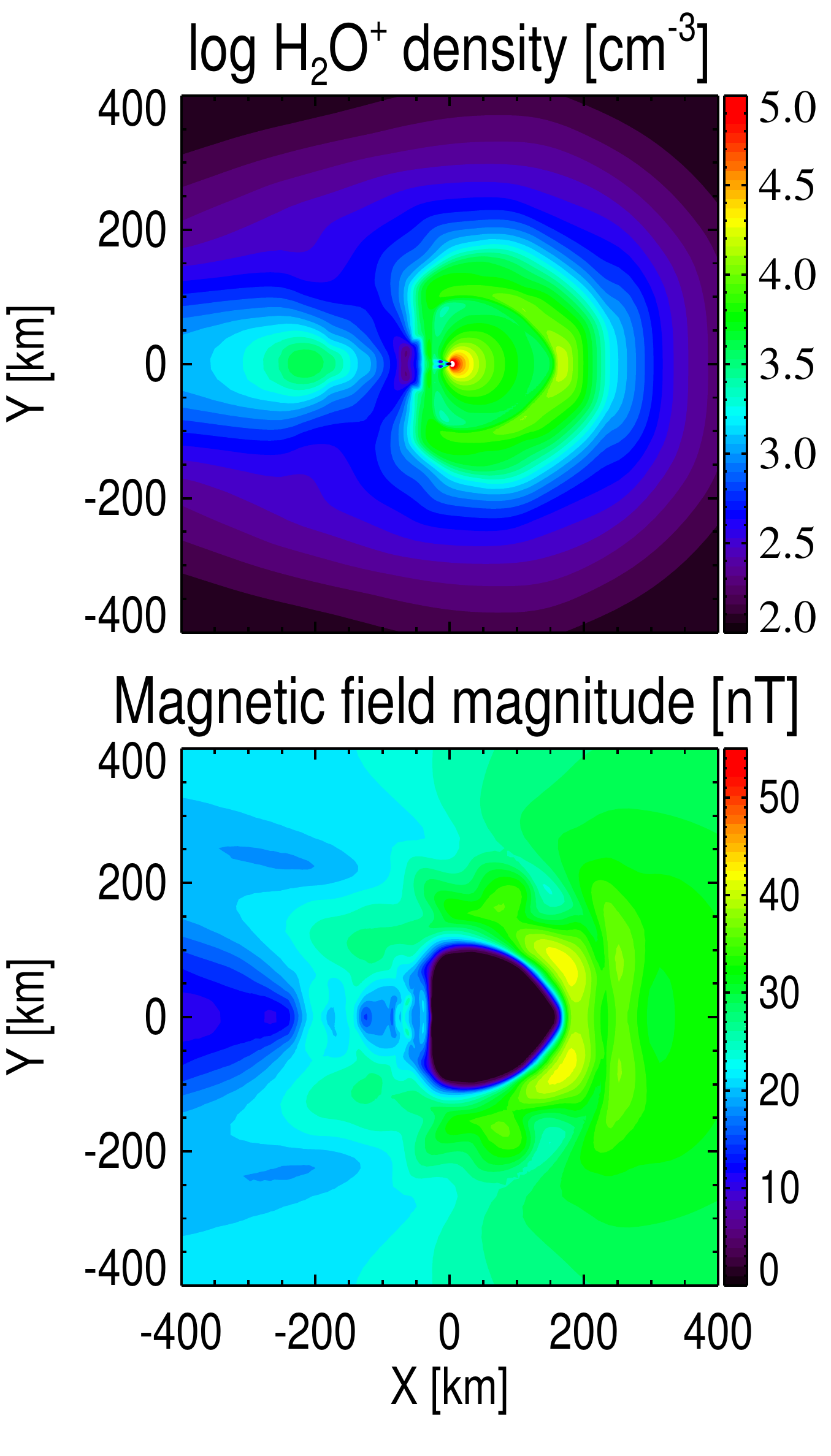} \\
\end{tabular}
\caption{The Hall MHD simulation results. The left columns plot the y=0 plane while the right columns 
plot the z=0 plane. The upper panels are for the cometary ion density while the lower panels show the magnetic field magnitude.}
\label{fig:hallyz}
\end{figure*}

\begin{figure*}
\begin{tabular}{cc}
\includegraphics[width=0.5\linewidth]{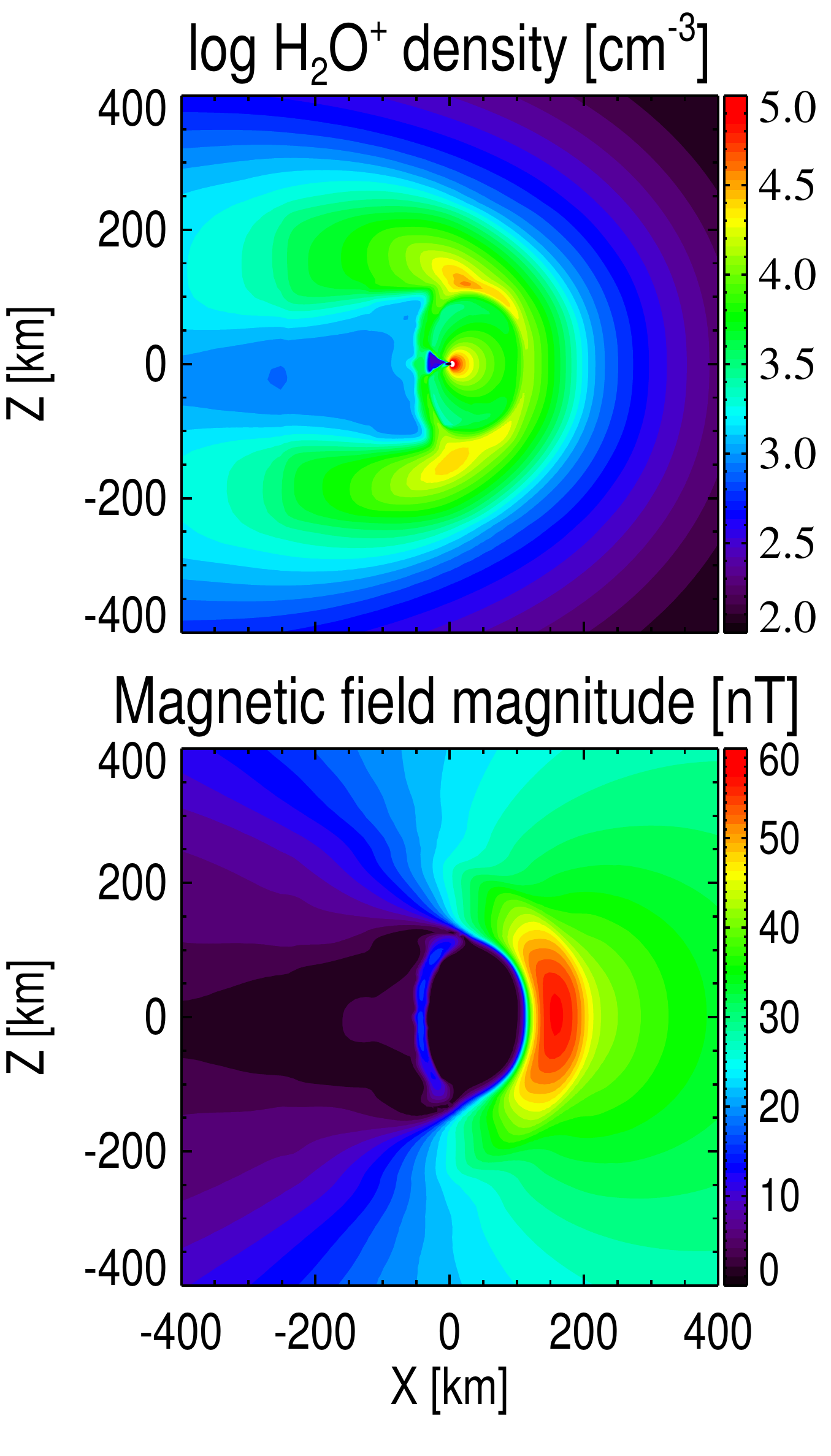} &
\includegraphics[width=0.5\linewidth]{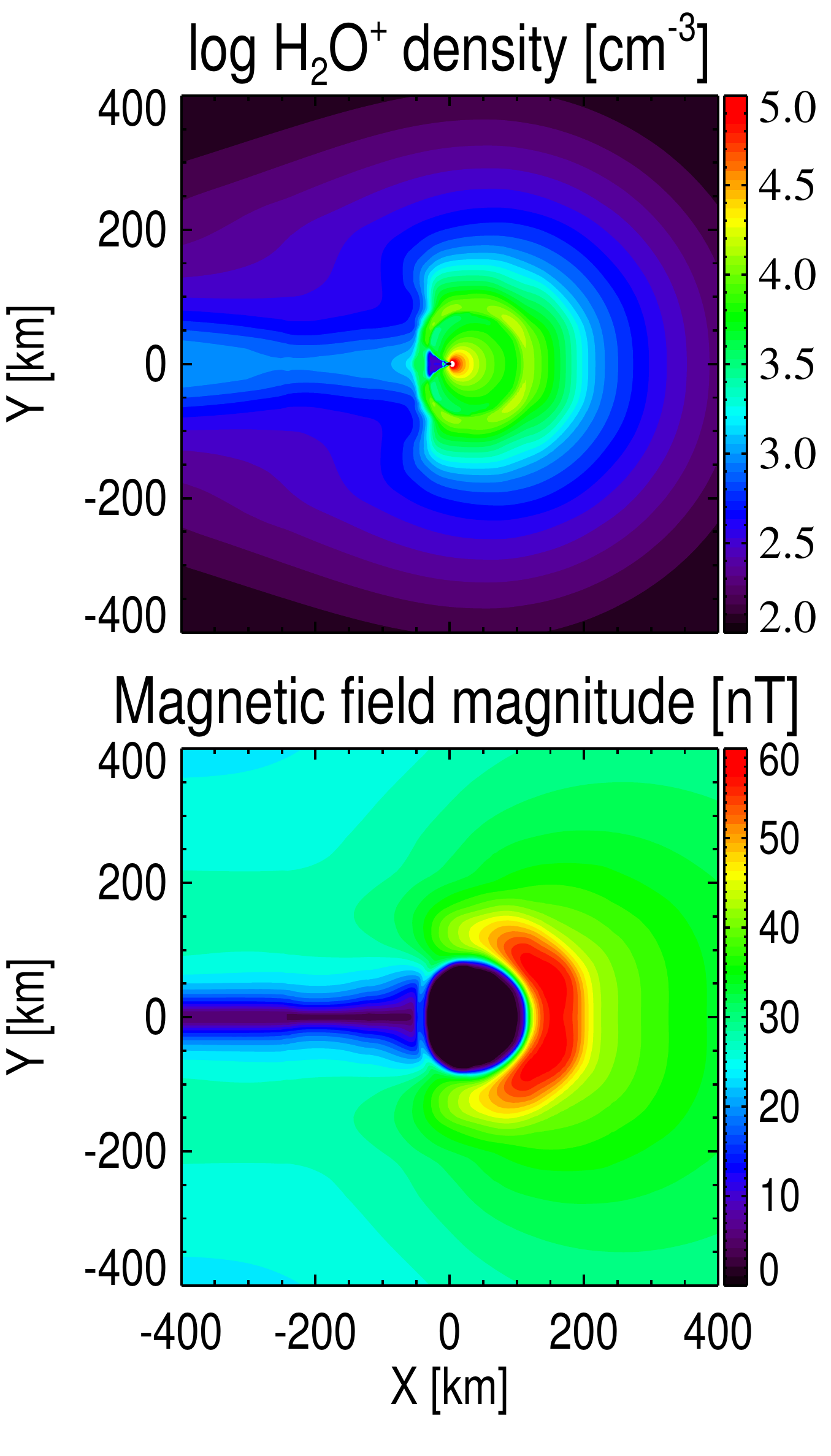} \\
\end{tabular}
\caption{A reproduction of the simulation results from \protect \cite{Huang_2016}. Multi-fluid MHD results without the Hall effect to be compared with Figure 1.}
\label{fig:case2}
\end{figure*}

Figure\,\ref{fig:hallyz} shows the multifluid Hall MHD simulation results in the inner coma region (within 400\,km of the nucleus) with
an illuminated sphere at $t=600\,\text{s}$.
While the simulation results preserve symmetries about the y axis in the z=0 plane, they show pronounced asymmetries in the y = 0 plane. 
As a comparison, we reproduce the multifluid
simulation results without the Hall effect in the same region with 
an illuminated sphere for the same input parameters (Case 2 in \cite{Huang_2016}) in Figure\,\ref{fig:case2}. In \cite{Huang_2016},
the size of the diamagnetic cavity and the location of 
the contact surface agreed well with previous 
MHD simulations \citep{Rubin_2015} and hybrid simulations \citep{Koenders_2015}.
However, when the Hall effect is introduced, the ion pile-up region (with light yellow color) in the upper panel is distorted in the y=0
plane and looks completely different from the upper panel in Figure\,\ref{fig:case2}, where the distribution is symmetric about the z axis.
Some surface wave structures, which might be associated with the Kelvin-Helmoltz (hereafter K-H) instabilities reported in \cite{Rubin_2012}, 
can also be found in the upper panel in Figure\,\ref{fig:hallyz}. 

The magnetic field topology in the bottom panels in Figure\,\ref{fig:hallyz} is completely different from the bottom panels in 
Figure\,\ref{fig:case2}. When the model does not include the Hall effect, 
the diamagnetic cavity (bottom panels in Figure\,\ref{fig:case2}) is an isolated region and the magnetic field pile-up region
is just upstream of the diamagnetic cavity. When the Hall effect is introduced, the magnetic field configuration becomes more complex and
besides the `global' diamagnetic cavity, regions of very weak magnetic field (less than 10\,nT) can also be found in the lower right corner in
the y=0 plane in the bottom left panel in Figure\,\ref{fig:hallyz}. We suggest that it is the $\mathbf{J} \times \mathbf{B}$ force discussed in the
next paragraph that changes the magnetic field configuration in the inner coma region.
The magnetic field pile-up region in the y=0 plane
is shifted and is not located in the same region as in Figure\,\ref{fig:case2}. In the z=0 plane (the bottom right panel
in both Figure\,\ref{fig:hallyz} and Figure\,\ref{fig:case2}), the diamagnetic cavity
looks more or less the same between the two simulations. The biggest difference lays in the magnetic pile-up region. 
In Figure\,\ref{fig:hallyz}, only two small magnetic field pile-up regions are found
outside the diamagnetic cavity while in Figure \ref{fig:case2}, the magnetic pile-up region is a single region.

\begin{figure}
\center
\includegraphics[width=0.6\linewidth]{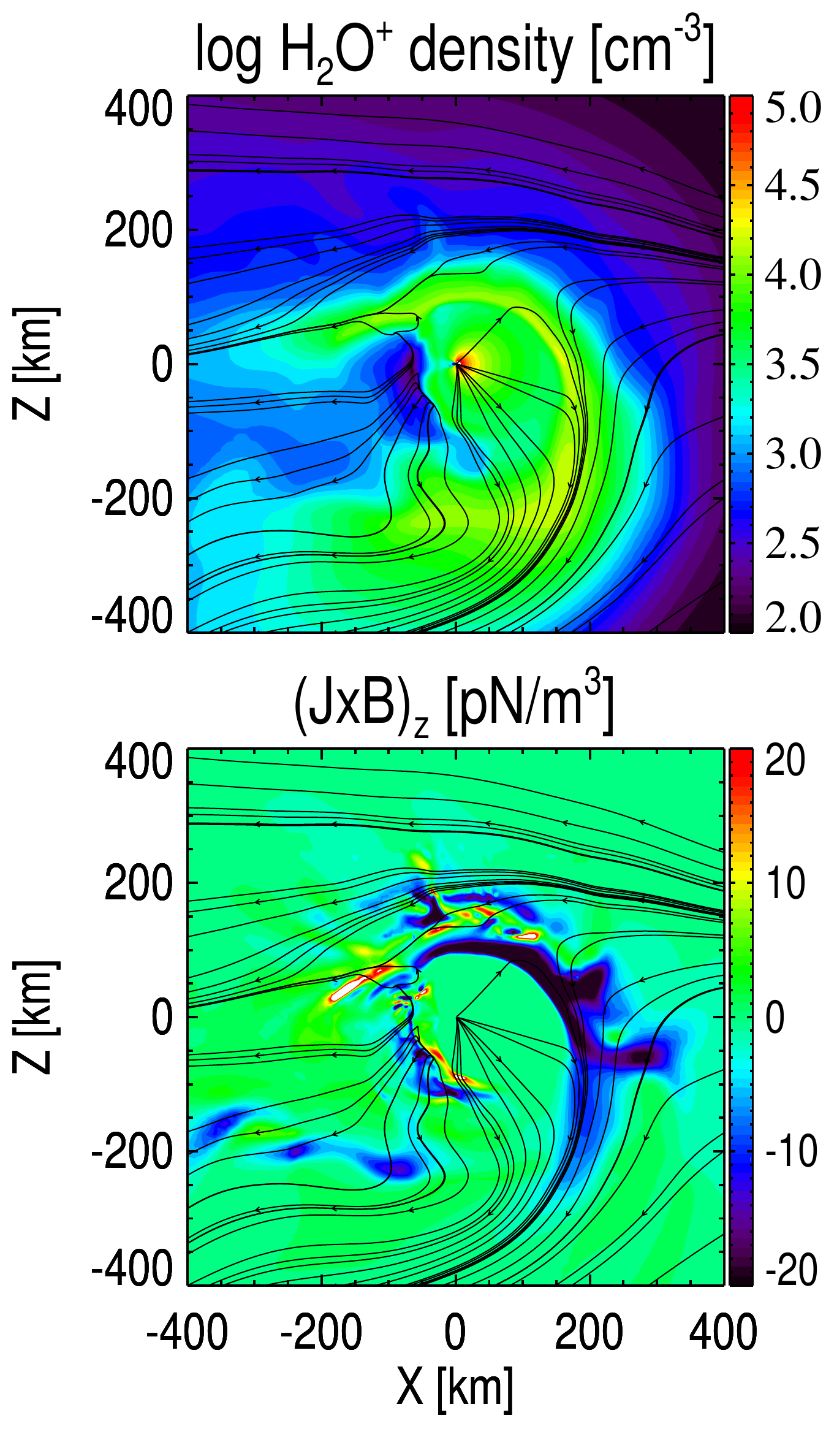} 
\caption{The upper panel shows the cometary ion density with their streamlines on the y=0 plane.
The bottom panels plots the z component of the 
$\mathbf{J} \times \mathbf{B}$ force density (in the unite of $\rm 1\times10^{-12}\,N/m^{3}$) on the y=0 plane.}
\label{fig:h2oprho_y}
\end{figure}

It is quite surprising that the simulated magnetic field changes so dramatically when the Hall effect is taken into account in the
induction equation. Besides, the Hall MHD simulation does not have a steady state solution despite the fixed upstream
solar wind conditions. The online movie (`inner\_coma\_movie.mp4') shows the evolution of the cometary ion density (with velocity streamlines) and the magnetic field 
between $t = 421$\,s and $t = 600$\,s. In the movie, the cometary ions move in the negative z direction. To illustrate this, we plot a 
snapshot of the cometary ion density with velocity streamlines at $t = 600$\,s in the upper panel of Figure\,\ref{fig:h2oprho_y}. 
This motion of the cometary ions can be explained by the $\mathbf{J} \times \mathbf{B}$ force, which is plotted in bottom panel of 
Figure\,\ref{fig:h2oprho_y}. This figure shows that along the global diamagnetic cavity boundary, 
the $\mathbf{J} \times \mathbf{B}$ force has a negative z component, which acts to move the cometary ions in the negative z direction.

\begin{figure}
\center
\includegraphics[width=0.95\linewidth]{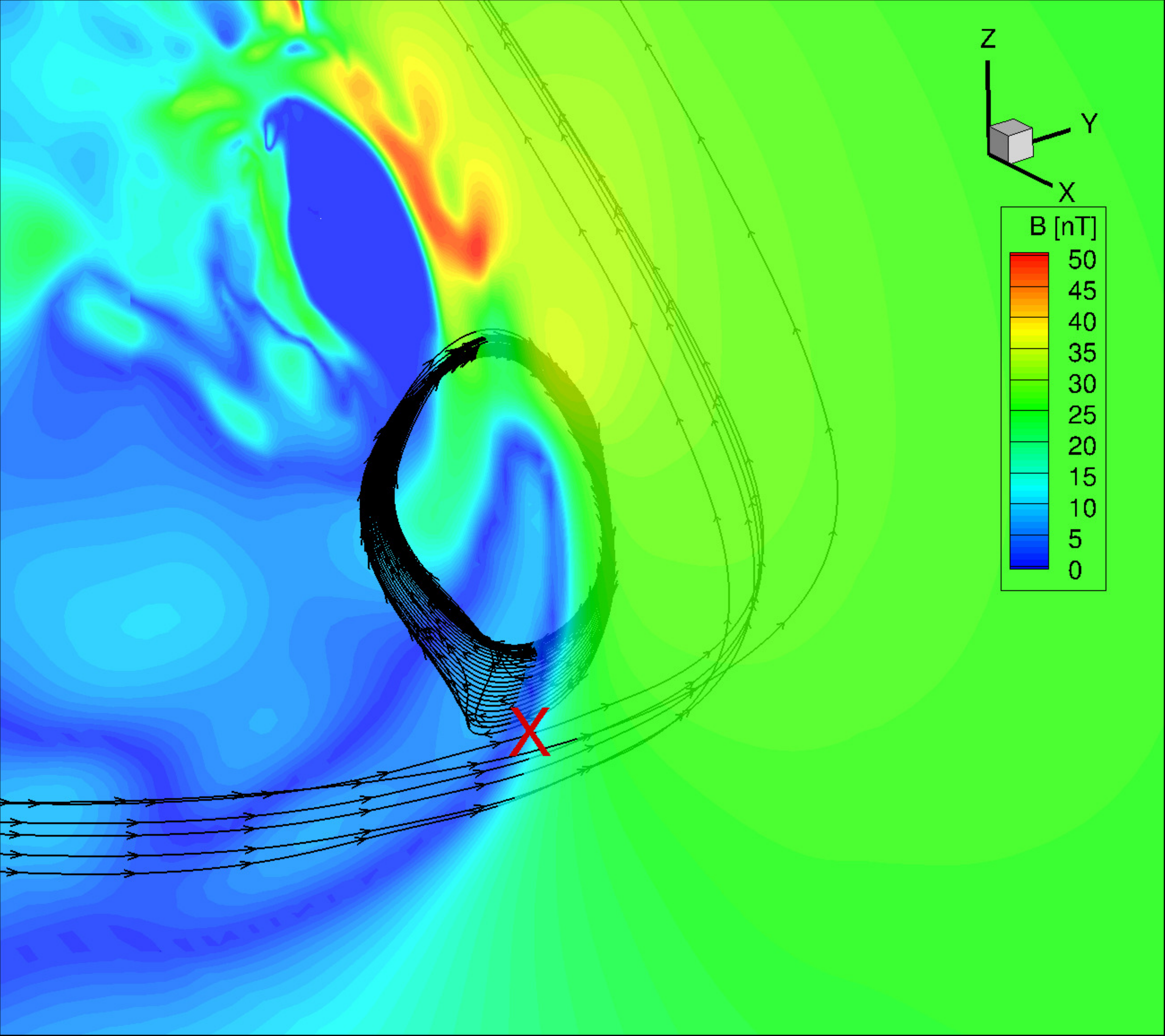} 
\caption{A 3D views of the magnetic field lines. The contour shows the magnetic field magnitude at the y=0 plane. Magnetic reconnection is
expected to occur near the red cross mark.}
\label{fig:3d}
\end{figure}

\begin{figure}
\center
\includegraphics[width=0.95\linewidth]{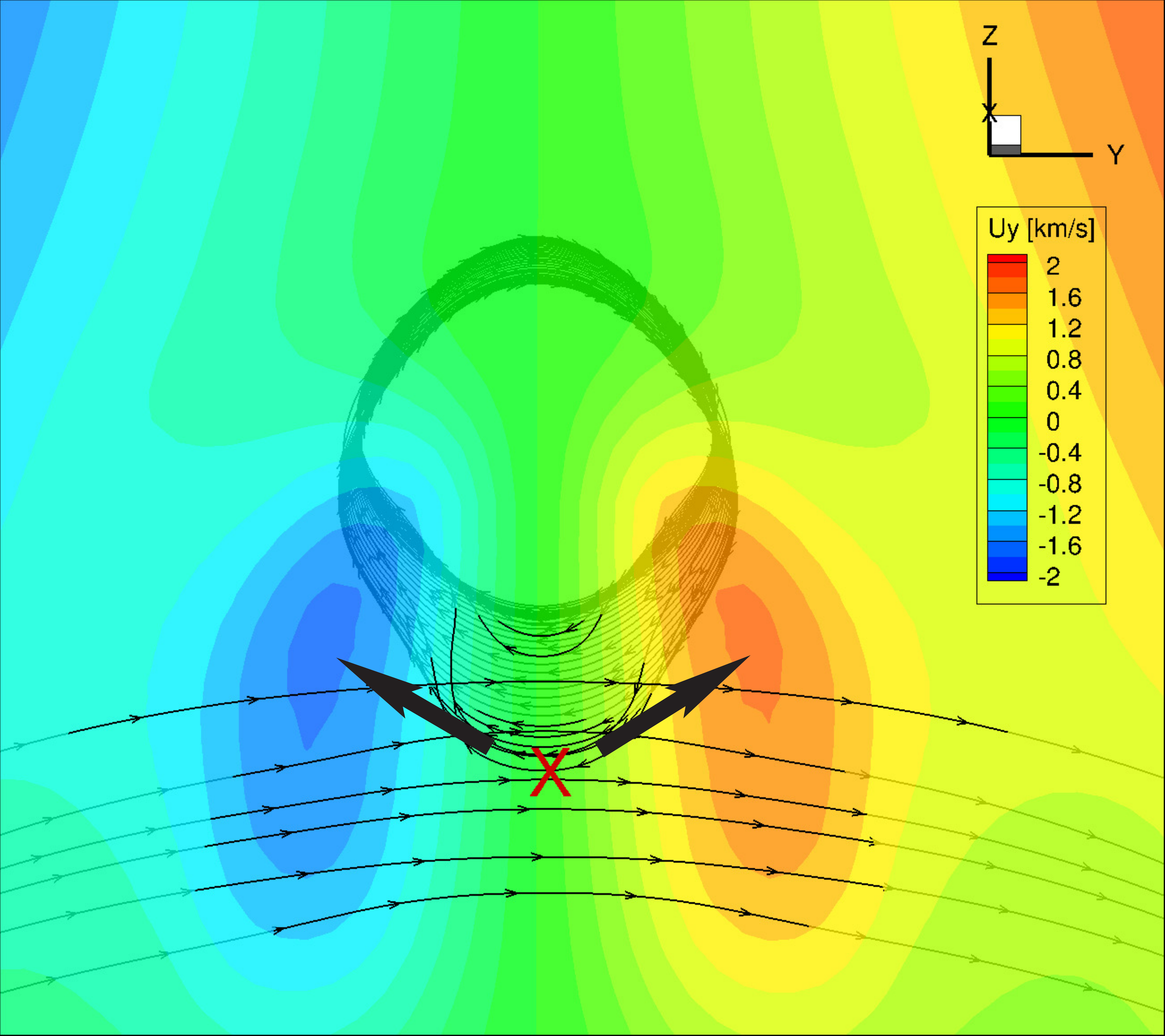} 
\caption{The $U_y$ component for the cometary ion velocity at the surface close to the magnetic reconnection surface, 
which is defined by 3 points: $[480,0,0]$\,km, $[500,20,33.5]$\,km and $[460,0,-33.5]$\,km. 
The cometary ions move to the +y direction on the right side while they move to the -y direction on the left side, 
near the magnetic null point denoted by the red cross mark, as indicated by the two black arrows.}
\label{fig:uy}
\end{figure}

Another important new observation from the Hall MHD simulation is the formation of the weak magnetic field regions in the lower right corner
in Figure \ref{fig:hallyz}. These structures appear as quasi-periodic structures in the online movie (`inner\_coma\_movie.mp4'). 
We provide a best estimate of the periods ranging from 10\,s to 50\,s, which are a combination of different harmonic periods,
based on the evolution of the magnetic pile-up regions in the online movie (`inner\_coma\_movie.mp4').
The periods depend on many factors, e.g., the plasma flow speed compared to the Alfv\'en speed, 
the strength of the currents as well as the direction and magnitude of the $\mathbf{J} \times \mathbf{B}$ force. It is impossible to  
calculate the exact periods in such a complex case.
To investigate how these weak magnetic field regions form,
we have examined the evolution of the magnetic field topology in 3D, which is animated in another two online movies (with different view angles, `reconnection\_movie\_view1.avi' and `reconnection\_moive\_view2.avi'). 
In the plasma, the magnetic field is approximately frozen into the electron fluid. 
Both the cometary ions in the coma and the electrons move in the
negative z direction (with the velocities separated by the currents). 
As the magnetic field moves with the electrons, the magnetic field is then draped in the negative z direction. 
A recent hybrid simulation by \cite{Koenders_2016} also showed
the draping signatures for comet CG at 2.0\,au. In the Hall MHD simulation, the draping of the magnetic field lines forms a configuration that 
favors magnetic reconnections. The online movies only animate 15\,s of the evolution (between $t = 425$\,s and 440\,s), but they clearly show
how the magnetic field reconnect and forms magnetic flux ropes. The magnetic reconnection reduces the magnetic field magnitude and
create the weak magnetic field regions. Figure\,\ref{fig:3d} plots the 3D magnetic field configuration. Magnetic reconnections are expected to
occur where the magnetic field lines bend strongly, denoted by an `X' mark on the figure. 
As magnetic reconnections occur, outflow is expected at the magnetic null point with opposite directions. Figure \ref{fig:uy} confirms that 
the plasma moves oppositely on the two sides. The outflow speed is close to the Alfv\'en speed near the reconnection regions, 
which is in the order of 1\,km/s.

\begin{figure}
\center
\includegraphics[width=0.6\linewidth]{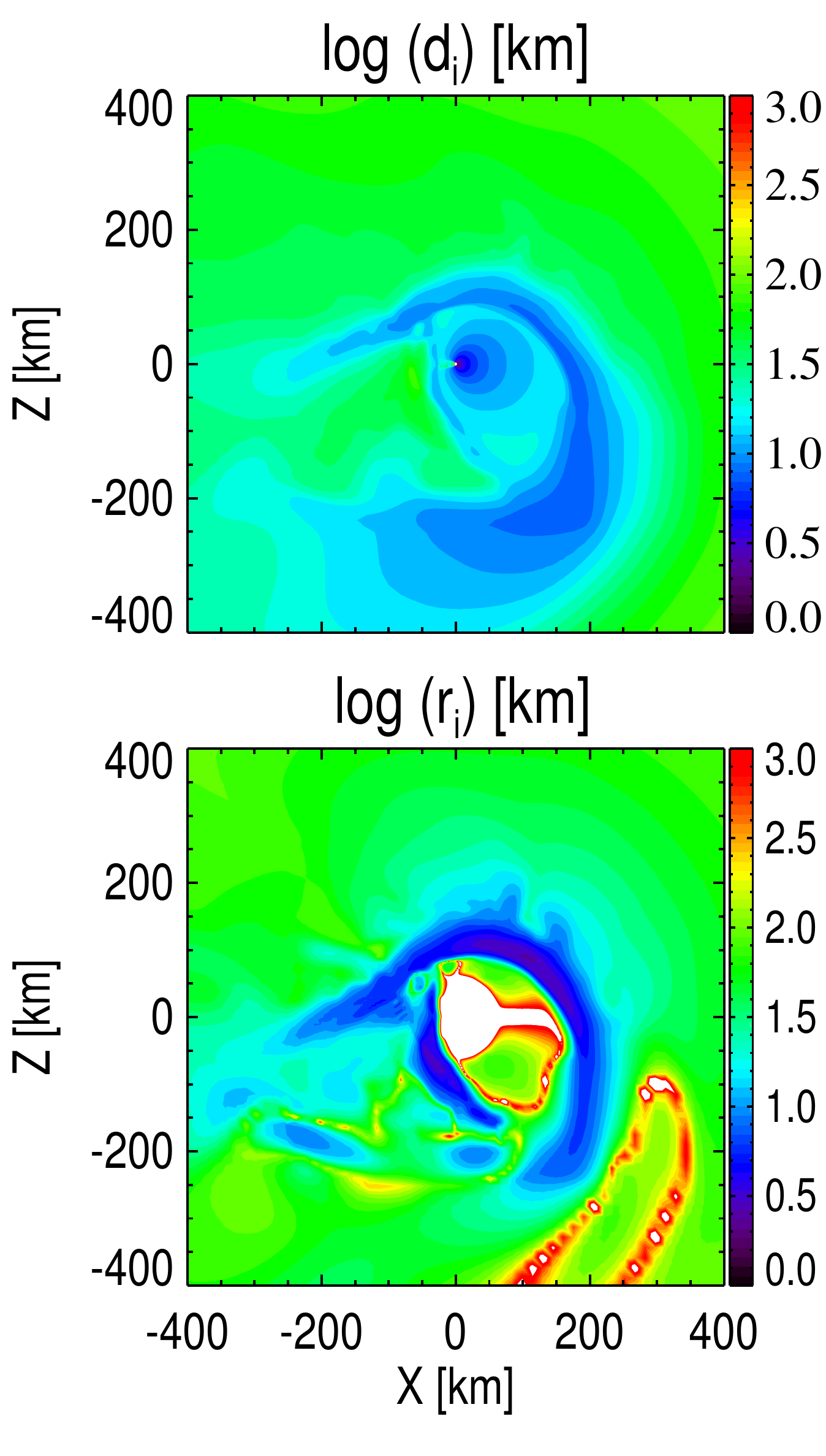}
\caption{The ion inertial length and the ion gyro radius for the cometary ion within 400\,km of the nucleus on the y=0 plane.}
\label{fig:di_ri}
\end{figure}

Why does the Hall effect matter in the inner coma of comet CG?
We argue that it is because the scale of the diamagnetic cavity is comparable to the ion inertial length ($d_i = \frac{m_i}{q_i} \sqrt{\frac{1}{\rho_i \mu_0}}$,
where $m_i$ is the ion mass, $q_i$ is the ion charge, $\rho_i$ is the ion mass density, $\mu_0$ is the magnetic permeability of vacuum)
and the ion gyro radius ($r_i = \frac{v_{th,i} m_i}{q_i B}$, where $v_{th,i}$ is the ion thermal speed, $B$ is the magnetic field magnitutde).
In the inner coma, the cometary ions dominate, so the ion inertial length and the gyro radius for the cometary ions are 
responsible for the physical processes. Figure \ref{fig:di_ri} plots the ion inertial length and the gyro radius
for the cometary ions. The ion inertial length is slightly less than 10\,km in the ion pile up region, and it is in the order of 
100\,km outside the global diamagnetic cavity and the ion pile-up region. The gyro radius is very large where the magnetic field is small.
Except in the weak magnetic field regions, the gyro radius has similar distributions as the ion inertial length.
As the size of the global diamagnetic cavity is about 100\,km, the ion inertial length and the ion gyro radius are not much smaller than
the global diamagnetic cavity. \cite{Dorelli_2015} showed that Hall currents within the magnetopause and magnetotail current sheets 
have a significant impact on the global structure of Ganymede's magnetosphere, 
because the magnetopause standoff distance is not much larger (order of 10) than the ion inertial length. In our case, the ratio of the size
of the diamagnetic cavity and the ion inertial length (or gyro radius) is in order of 10 or less. We put forward an argument that if the
ion inertial length (or gyro radius) is not much smaller than the characteristic length of the magnetosphere 
(the diamagnetic cavity in our case, the magnetopause standoff distance in Ganymede's magnetosphere), Hall MHD simulations are
necessary to capture the correct global structure of the magnetosphere. 

\section{Summary and Discussions}
In this work, we performed a multi-fluid Hall MHD simulation to study the cometary plasma environment in the 
inner coma region of comet 67P/Churyumov-Gerasimenko.  With the same model set up as \cite{Huang_2016}, 
the Hall MHD simulation shows a very different picture: the inner coma is no longer symmetric 
and low magnetic field regions can form outside the global diamagnetic cavity and the solution is time-dependent.

The only difference between the Hall MHD simulation and the classical MHD simulations by \cite{Huang_2016} is that
the Hall velocity term is considered in the magnetic induction equation, which means that
the current can affect the evolution of the magnetic field. It is well known that the Hall effect is important in
magnetic reconnections \citep{Birn_2001}, partially due to relative weak magnetic field and strong currents 
near the magnetic null point. Hall MHD simulations of the magnetospheres of planets and moons typically do not show significant 
differences compared to the classical MHD simulations except in the regions where magnetic reconnections occur like
the dayside magnetopause or the nightside magnetotail. \cite{Dorelli_2015} reported that Hall effect is
important in Ganymede's magnetosphere because the magnetopause standoff distance is in the order of 10 times larger
than the ion inertial length. One would not expect the Hall MHD simulations dramatically to change
the simulated inner coma environment for a comet because magnetic field reconnections have only been reported on the 
nightside \citep{Huang_2016}, but our simulations show that in fact the results change dramatically.

The diamagnetic cavity is a unique feature in the cometary environment, which is not shared by other planets or moons in the solar system,
and it has received lots of attention since the Giotto mission 
\citep{Neubauer_1986, Cravens_1986, Goetz_2016, Goetz_2016_2, Huang_2016MNRAS, Madanian_2017}.
However, it has not been realized that currents along the diamagnetic cavity boundary may 
change the global structure of the inner coma. In the comet CG case, as the ion inertial length (or gyro radius) is not much smaller than the size
of the global diamagnetic cavity, the Hall effect plays an important role in the evolution of the cometary 
plasma environment in the inner coma region, which is confirmed by our Hall MHD simulation.
The situation might be different for a much more active comet. For example, the size of the diamagnetic cavity for comet 1P/Halley is
about 4500\,km \citep{Neubauer_1986, Cravens_1986}, which will need to be compared with the ion inertial length (or gyro radius)
to see whether the Hall effect is important there.

The most important feature from the Hall MHD simulations is that there can be dayside magnetic reconnection, which can create weak 
magnetic field regions outside the global diamagnetic cavity. One of the most puzzling observations from the 
Rosetta Plasma Consortium (RPC)
is that the magnetometer observed weak magnetic field at a distance much farther away than the predicted diamagnetic cavity.
\cite{Goetz_2016, Goetz_2016_2} explained the weak magnetic field observations as K-H instabilities propagating along the cavity boundary
and \cite{Huang_2016MNRAS} explained them as short-lived enhanced electron pressure along magnetic field lines.
The Hall MHD simulation may provide a third option, magnetic field reconnection on the dayside.
Further investigation and data comparison is necessary, but at this point, we refer this to future studies.

\section*{Acknowledgements}

This work was supported by contracts Jet Propulsion Laboratory no. 1266313 and no. 1266314 from the US Rosetta Project and NASA grant NNX14AG84G from the Planetary Atmospheres Program.

The authors would like to thank the ROSINA team for supporting this research. The authors also thank the ESA Rosetta team for providing the opportunities to study this unique comet and their continuous support.

The authors would like to acknowledge the following high-performance computing resources: 
Yellowstone (ark:/85065/d7wd3xhc),
provided by NCAR's Computational and Information Systems Laboratory, sponsored by the National Science Foundation;
Pleiades, provided by the NASA Supercomputer Division at Ames; and 
Extreme Science and Engineering Discovery Environment (XSEDE), 
supported by National Science Foundation grant number ACI-1053575

%%%%%%%%%%%%%%%%%%%%%%%%%%%%%%%%%%%%%%%%%%%%%%%%%%



\bibliographystyle{mnras}
\bibliography{reference} % if your bibtex file is called example.bib

%%%%%%%%%%%%%%%%%%%%%%%%%%%%%%%%%%%%%%%%%%%%%%%%%%

% Don't change these lines
\end{document}